\documentclass[aps,prd,twocolumn,showpacs,floatfix,superscriptaddress,preprintnumbers]{revtex4}
\usepackage{amssymb}
\usepackage{epsfig}
\usepackage{graphicx}
\usepackage{stmaryrd}

\def\10{$SO(10)$}
\def\21{SU(2) $\otimes$ U(1) }

\def\422{$SU(4) \otimes SU(2) \otimes SU(2)$}
\def\321{SU(3) $\otimes$ SU(2) $\otimes$ U(1)}

\def\lsim{\raise0.3ex\hbox{$\;<$\kern-0.75em\raise-1.1ex\hbox{$\sim\;$}}}
\def\gsim{\raise0.3ex\hbox{$\;>$\kern-0.75em\raise-1.1ex\hbox{$\sim\;$}}}
\def\vev#1{\left\langle #1\right\rangle}

\newcommand{\AddrAHEP}{%
  AHEP Group, Institut de F\'{\i}sica Corpuscular --
  C.S.I.C./Universitat de Val{\`e}ncia \\
  Edificio Institutos de Paterna, Apt 22085, E--46071 Valencia, Spain}

 \newcommand{\ba}{\begin{array}}
\newcommand{\ea}{\end{array}}
\relax
\def\321{$SU(3)\times SU(2)\times U(1)$}
\def\mnu{{\cal M}_{\nu }}
\newcommand{\Sol}  {\textrm{sol}}

\newcommand{\Atm}  {\textrm{atm}}

\newcommand{\Dms}  {\Delta m^2_\Sol}
\newcommand{\Dma}  {\Delta m^2_\Atm}

\begin{document}
\preprint{IFIC/07-13}
\renewcommand{\Huge}{\Large}
\renewcommand{\LARGE}{\Large}
\renewcommand{\Large}{\large}
\title{Predictive flavour symmetries of the neutrino mass matrix}
\author{M.~Hirsch} \email{mahirsch@ific.uv.es} \affiliation{\AddrAHEP}
\author{Anjan S.  Joshipura} \email{Anjan.Joshipura@ific.uv.es}
\affiliation{\AddrAHEP} \affiliation{Theoretical Physics Group,
  Physical Research Laboratory, Navrangpura, Ahmedabad 380 009, India}
\author{S. Kaneko} \email{satoru@ific.uv.es}\affiliation{\AddrAHEP}
\author{J.~W.~F.~Valle} \email{valle@ific.uv.es}
\affiliation{\AddrAHEP}

\date{\today}

\begin{abstract}

  Here we propose an $A_4$ flavour symmetry model which implies a
  lower bound on the neutrinoless double beta decay rate,
  corresponding to an effective mass parameter $M_{ee} \gsim 0.03$~eV,
  and a direct correlation between the expected magnitude of CP
  violation in neutrino oscillations and the value of
  $\sin^2\theta_{13}$, as well as a nearly maximal CP phase $\delta$.

\end{abstract}

\pacs{
11.30.Hv       
14.60.-z       
14.60.Pq       
14.80.Cp       
}

\maketitle


Unless flavour symmetries are assumed, particle masses and mixings are
generally undetermined in gauge theories.  Understanding mass and
mixing constitutes one of the biggest challenges in elementary particle
physics. Current observations do not determine all elements of the
effective neutrino mass matrix $\mnu$ completely and this will be a
great challenge even for future experiments.  Therefore theoretical
ideas restricting the structure of $\mnu$ are needed in order to guide
future searches.  One such input studied extensively is the assumption
that some entries in the neutrino mass matrix
vanish~\cite{Frampton:2002yf}.
While the phenomenological implications of the assumed zeros in the
texture of $\mnu$ are straightforward to derive~\cite{Dev:2006qe}, it
is a non-trivial task to produce a good symmetry leading to such zeros
and a diagonal charged lepton mass matrix simultaneously. Although for
any desired texture structure of the mass matrices such a symmetry is
in principle always present, this symmetry and the associated Higgs
content are sometimes discouragingly complex~\cite{Grimus:2004hf}.

Here we propose a predictive flavour symmetry for leptons based on a
relatively small and simple flavour group, namely $A_4$ or its $Z_3$
subgroup, and briefly analyse its phenomenological implications.
We show how this provides a simple means of understanding some of the
two-zero textures of $\mnu$ studied earlier~\cite{Dev:2006qe}.

The discrete group $A_4$ is a 12 element group consisting of even
permutations among four objects. The group is small enough to lead to
a simple model but large enough to give interesting predictions. The
distinguishing feature of $A_4$ compared to other smaller discrete
groups is the presence of a 3 dimensional irreducible representation
appropriate to describe the three generations. This has been exploited
in a number of variants. Originally, the $A_4$ was proposed
\cite{Ma:2001dn,Babu:2002dz} for understanding degenerate neutrino
spectrum with nearly maximal atmospheric neutrino mixing angle.  More
recently, predictions for the solar neutrino mixing angle have also
been incorporated in so-called tri-bi-maximal~\cite{Harrison:2002er}
neutrino mixing
schemes~\cite{King:2006np,Ma:1,He:2006dk,Hirsch:2005mc,Altarelli:2005yx,Ma:2004zv}.
There also exist attempts at unified $A_4$ models~\cite{Ma:2006wm}.
The resulting models however are not always simple and usually require
many Higgs fields. Here we show that a very simple model based on
$A_4$ leads to two-zero textures for $\mnu$.

The $L_i$ are assigned to the triplet representation in all the $A_4$
models proposed so far.  Here we propose the opposite assignment
indicated in Table~\ref{tab:Multiplet}, where the $L_i$ are assigned
to the $1,1',1''$ representations.
\begin{table}
   \centering
\begin{math}
\begin{array}{|c|c|c|c|c|c|c|c|} \hline
& L_1 & L_2 & L_3 & l_{Ri} & \nu_{Ri} & \Phi_i  & \Delta\\
\hline
SU(2) & 2 & 2 & 2 & 1 & 1 & 2 & 3
\\
\hline
U(1) & -1 & -1 & -1 & -2 & 0 & 1 & 2
\\
\hline
A_4 & 1 & 1' & 1'' & 3 & 3 & 3 & 1'\ {\rm or}\  1''
\\
\hline
\end{array}
\end{math}
\caption{Lepton multiplet structure of the model}
\label{tab:Multiplet}
\end{table}
The $l_{Ri}$ as well as the Higgs doublets responsible for lepton
masses transform as $A_4$ triplets, while the (undisplayed) quarks and
the SU(2) Higgs doublet that gives their masses are all singlets under
$A_4$.  This leads to the following terms responsible for the 
lepton masses:
\begin{eqnarray}
\label{eq:yukawa1}
 -{\cal L}&=&h_1 \bar{L}_1(l_R\Phi)_1+h_2
\bar{L}_2(l_R\Phi)_1'+h_3 \bar{L}_3(l_R\Phi)_1'' \nonumber\\
&+&h_{1D} \bar{L}_1(\nu_R\Phi)_1+h_{2D}
\bar{L}_2(\nu_{R}\Phi)_1'+h_{3D} \bar{L}_3(\nu_R\Phi)_1'' \nonumber\\
&+&\frac{M}{2}\nu_{Ri}^TC\nu_{Ri}+{\rm H.c.}~,
\end{eqnarray} 
where the quantities in parenthesis denote products of two
$A_4$-triplets $l_R$ (or $\nu_R$) and $\Phi$ forming the
representations $1,1',1''$ respectively.  Note that
Eq.~(\ref{eq:yukawa1}) includes the most general terms allowed by the
symmetry and field content in Table~\ref{tab:Multiplet}. Hence, in
contrast to many other $A_4$ models, here one does not need to impose
any additional symmetry to forbid unwanted terms.

Earlier studies on $A_4$ have shown that it is possible to obtain a
minimum of the Higgs potential with equal vacuum expectation values
(vevs)~\cite{Ma:2001dn}
\begin{equation}
\label{eq:vev}
\vev{\Phi_1^0}=
\vev{\Phi_2^0}=
\vev{\Phi_3^0} \equiv \frac{v}{\sqrt{3}}~.
\end{equation} 
This minimum leads to charged lepton and Dirac neutrino mass matrices
$M_l$ and $m_D$ given by, respectively
\begin{eqnarray*}
\label{eq:mlmd} 
M_l = v~{\rm
  diag}(h_1,h_2,h_3)U\\
m_D=v~{\rm  diag}(h_{1D},h_{2D},h_{3D})U~,
\end{eqnarray*}
with 
\begin{equation} 
\label{eq:u} 
U=\frac{1}{\sqrt{3}}\left( \begin{array}{ccc}
    1&1&1\\
    1&\omega&\omega^2\\
    1&\omega^2&\omega\\ \end{array} \right)~,~~~~\omega\equiv e^{\frac{2 \pi i}{3}}~.
\end{equation} 
The above $M_l,m_D$ imply that the symmetry basis $L_i$ also
corresponds to the mass basis and only the right handed fields need to
be redefined.  As a result, the neutrino mass matrix following from
eqs.(\ref{eq:yukawa1},~\ref{eq:vev}) after the seesaw
diagonalization~\cite{Valle:2006vb} is already in the flavour basis
and is given by
\begin{equation} 
\label{eq:type1} 
{\cal M}_{\nu f}^I=m_DM_R^{-1}m_D^T=
\frac{v^2}{M}\left(\begin{array}{ccc}
  h_{1D}^2&0&0\\
  0&0&h_{2D}h_{3D}\\
  0&h_{2D}h_{3D}&0\\ \end{array} \right)~. 
\end{equation} 
This has the same zero textures as obtained in \cite{Babu:2002dz}
except that only two (instead of three) neutrinos are degenerate.  As
noted in \cite{Babu:2002dz}, this texture by itself is not complete
and one needs to modify it. For example, one can supersymmetrize the
above scenario and use radiative corrections to split the degeneracy
and obtain predictions for the mixing angles and masses as
in~\cite{Babu:2002dz}.

Here we choose a different approach, introducing a triplet field
$\Delta$~\cite{schechter:1980gr} transforming either as a 1'' or as a
1' under $A_4$, as in Table~\ref{tab:Multiplet}.
In the first case a small induced vev $\vev{\Delta^0}\equiv u$ for its
neutral component leads to a type-II neutrino mass matrix contribution
given as
\begin{equation}
\label{eq:type2} {\cal M}_{\nu}^{II}=\left( \begin{array}{ccc}
  0&\lambda u&0\\
  \lambda u&0&0\\
  0&0&\lambda'u\\ \end{array} \right) ~,
\end{equation} 
where $\lambda,\lambda'$ are two Yukawa couplings (another hybrid
model based on $A_4$ and using both type-I and type-II contributions
to neutrino masses has been considered in \cite{Chen:2005jm}).  The
total neutrino mass matrix is given by the sum of eq.(\ref{eq:type1})
and (\ref{eq:type2}) and has the form
\begin{equation} 
\label{eq:mnu1} \mnu=\left( \begin{array}{ccc}
  a&x&0\\
  x&0&b\\
  0&b&y\\ \end{array} \right) ~,
\end{equation} 
where $a,b$ and $x,y$ refer to the type-I and type-II contributions,
respectively. The above arguments provide a simple derivation of the
two-zero texture classified as $B_1$ in Ref.~\cite{Frampton:2002yf}).

Alternatively, had the triplet been assigned to the $1'$
representation of $A_4$ then we would have obtained 
\begin{equation} 
\label{eq:mnu2} \mnu=\left( \begin{array}{ccc}
    a&0&x\\
    0&y&b\\
    x&b&0\\ \end{array} \right)~, 
\end{equation} 
a texture classified as $B_2$ in \cite{Frampton:2002yf}. It is
possible to modify the assignment of various $L_i$ fields among
different singlet representations of $A_4$.  This either results in
one of the two above textures, or in a texture which is not viable
phenomenologically.  Thus the realization of the the $A_4$ flavour
symmetry proposed here leads to just two viable two-zero textures,
which are quite predictive as we will show.

The vacuum structure given in eq.(\ref{eq:vev}) breaks the $A_4$
preserving a $Z_3$ subgroup~\cite{He:2006dk}.  This $Z_3$ in our case
gets broken spontaneously by the triplet vacuum expectation value.
Thus the type-I contribution in eq.(\ref{eq:type1}) is $Z_3$ invariant.
Interestingly, the converse and more powerful statement is also true.
One can argue the above two-zero texture to be a consequence of the
(spontaneously broken) $Z_3$ symmetry instead of the full $A_4$ as we
now show.

The $Z_3$ group under consideration is generated by
$(1,z,z^2)~,~z^3=1$ with the leptons transforming as 
\begin{eqnarray} 
\label{eq:z3}
L_i&\rightarrow& Z_{ij}^LL_j~,\nonumber\\
(l_{Ri},\nu_{Ri})& \to & Z_{ij}^R(l_{Rj},\nu_{Rj})~,
\end{eqnarray}
where $Z^L={\rm diag}~(1,\omega,\omega^2)$ and
\begin{equation} \label{zr}
Z^R=\left( \begin{array}{ccc}
0&0&1\\
1&0&0\\
0&1&0\\ \end{array} \right) ~.
\end{equation}
Note that the fields which earlier transformed as triplets under $A_4$
are now put into a reducible representation of the $Z_3$ group.
Let us now demand that $M_l,m_D$ and $M_R$ are invariant under 
the above defined $Z_3$. This implies
\begin{equation} \label{invariance}
Z^{L\dagger}M_lZ^R=M_l~;~Z^{L\dagger}m_DZ^R=m_D~;~Z_R^TM_RZ^R=M_R~.
\end{equation}
It is straightforward to show that the above invariance implies that
both $M_l,m_D$ must have the form
\begin{equation}
\label{eq:form}\left(\begin{array}{ccc}
X&X&X\\
A&\omega A& \omega^2 A\\
B&\omega^2 B& \omega B\\
\end{array}\right)\end{equation}
The above form coincides with that obtained in eq.~(\ref{eq:mlmd})
with proper identification of parameters. The right handed neutrino
mass matrix now has the following general form~\footnote{This more
  general form would arise in a model containing an $SU(2)_L$-singlet
  but $A_4$ triplet Higgs $\eta$ with a $Z_3$-preserving vev.}
\begin{equation} \label{mr}\left( \begin{array}{ccc}
M_1&M_2&M_2\\
M_2&M_1&M_2\\
M_2&M_2&M_1\\ \end{array} \right)
\end{equation}
In spite of this more complicated form, it is easy to see that the
type-I contribution has exactly the same zero texture as in
eq.(\ref{eq:type1}) 
which is therefore more general than its derivation through the seesaw
mechanism used here. It simply follows from the $Z_3$ invariance of the
effective neutrino mass matrix:
\begin{equation} Z^{LT}\mnu Z^L=\mnu 
\end{equation}
irrespective of the underlying dynamics. For example, the same form
would arise in a model without the right handed neutrinos but
containing a $Z_3$-singlet Higgs triplet with a non-zero vev. As in
the $A_4$ case, one can introduce a triplet $H$ transforming as
$\omega^2$ under $z$, and whose vev will now break $Z_3$ to give the
required two-zero texture as in eq.~(\ref{eq:type2}).
 
We now turn to the phenomenological implications.  
The main feature of two-zero texture models, such as the ones derived
here, is their power in predicting the as yet undetermined neutrino
parameters. Current neutrino oscillation experiments determine two
mass splittings $\Dma$ and $\Dms$ and the corresponding mixing angles
$\theta_{12}$ and $\theta_{23}$, with some sensitivity on
$\theta_{13}$ which is bounded~\cite{Maltoni:2004ei}.  The Dirac CP
phase will be probed in future oscillation experiments. Similarly, the
absolute neutrino mass scale will be probed by future cosmological
observations~\cite{Lesgourgues:2006nd}, tritium beta
decays~\cite{Drexlin:2005zt} and neutrinoless double beta decay
experiments~\cite{Hirsch:2006tt} with improved sensitivity. The latter
will also shed light on the two Majorana CP phases which are hard to
test otherwise, as they do not affect lepton number conserving
processes.
The general $3\times 3$ light neutrino mass matrix $\mnu$ in the
flavour basis contains \emph{a priori} nine independent real
parameters, once the three unphysical phases associated with the
charged lepton fields are removed. 
In contrast, in the proposed model all the above nine parameters are
given in terms of only five unknowns.  Hence the number of physical
parameters characterizing the charged current weak interaction is
reduced with respect to what is expected in the general
case~\cite{schechter:1980gr}.

\begin{figure}[t]
  \begin{center}
  \includegraphics[height=4.5cm,width=0.45\textwidth]{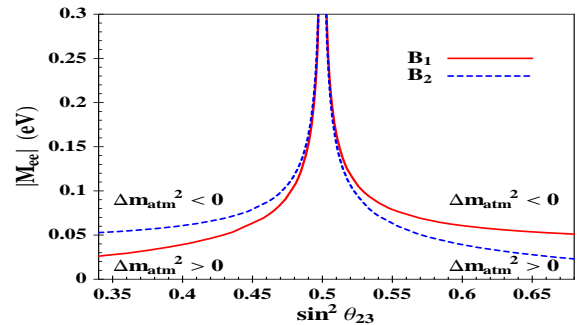}
  \end{center}
  \caption{Lower bound on neutrinoless double beta decay.}
  \label{fig:mee}
\end{figure}

We now illustrate these predictions. We first consider the mass
parameter characterizing neutrinoless double beta decay $|M_{ee}|$ 
which depends mainly on  $\theta_{23}$, as illustrated
in Fig.~\ref{fig:mee}. 
A remarkable feature of our $A_4$ flavour symmetry model is that it
implies the lower bound $|M_{ee}| \gsim 0.03$~eV, as seen in
Fig.~\ref{fig:mee}. 
This prediction correlates with the maximality of the atmospheric
mixing angle and lies within the range of planned experiments.  The
bound hardly depends on other parameters. For example, in contrast to
Ref.~\cite{Hirsch:2005mc}, it shows no strong dependence with the
value of the relevant Majorana phase. This follows from the more
stringent lower bound on the lightest neutrino mass obtained in the
present model. We note that $|M_{ee}|$ has, however, some dependence
on the value of $\Dma$ and the bound corresponds to $\Dma = 2 \times
10^{-3}$ eV$^2$.

We now turn to the predictions for CP violation and the parameter
$\delta$. 
\begin{figure}[h]
\begin{center}
\includegraphics[height=3.5cm,width=0.2365\textwidth]{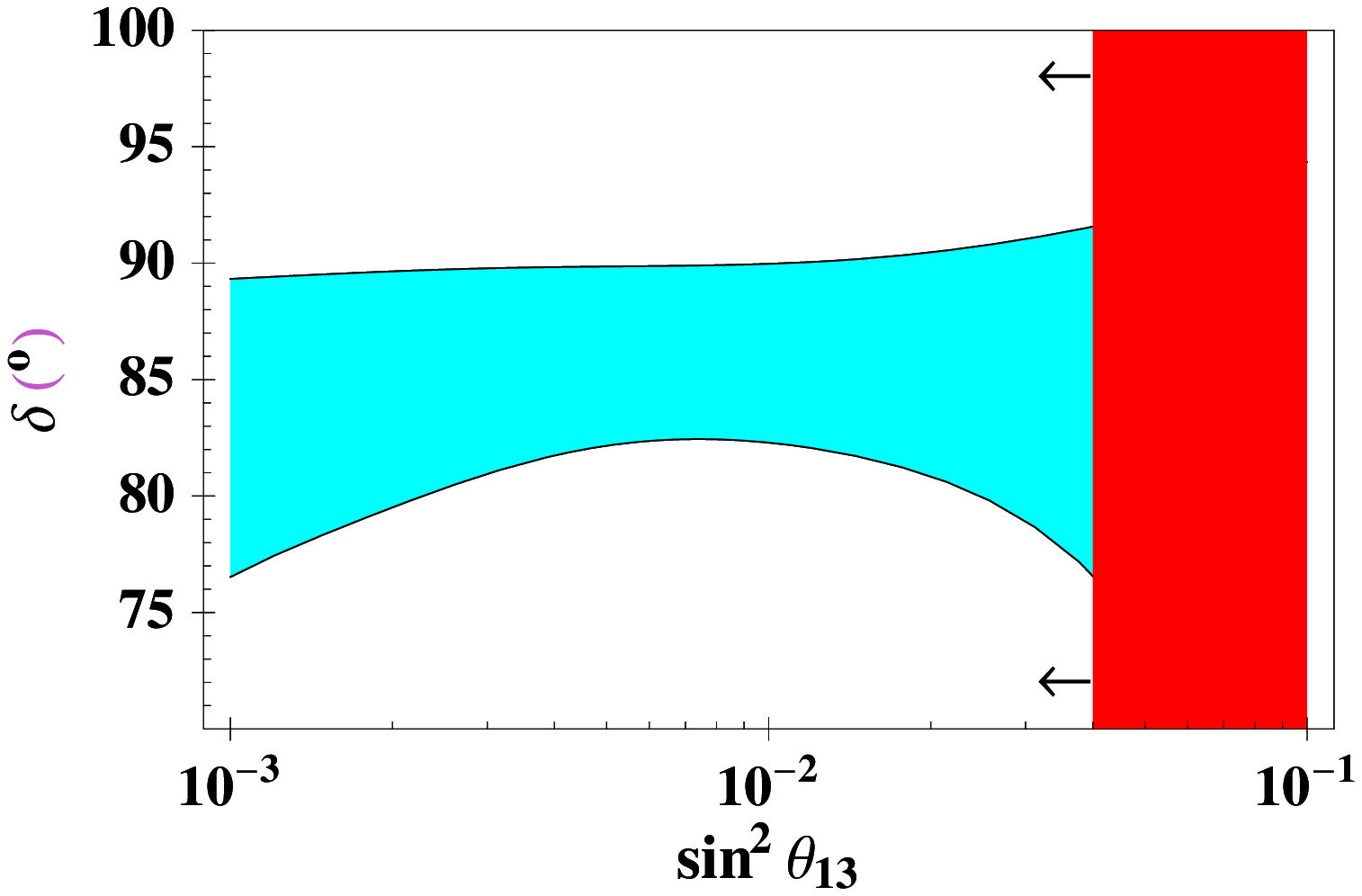} 
\includegraphics[height=3.5cm,width=0.2365\textwidth]{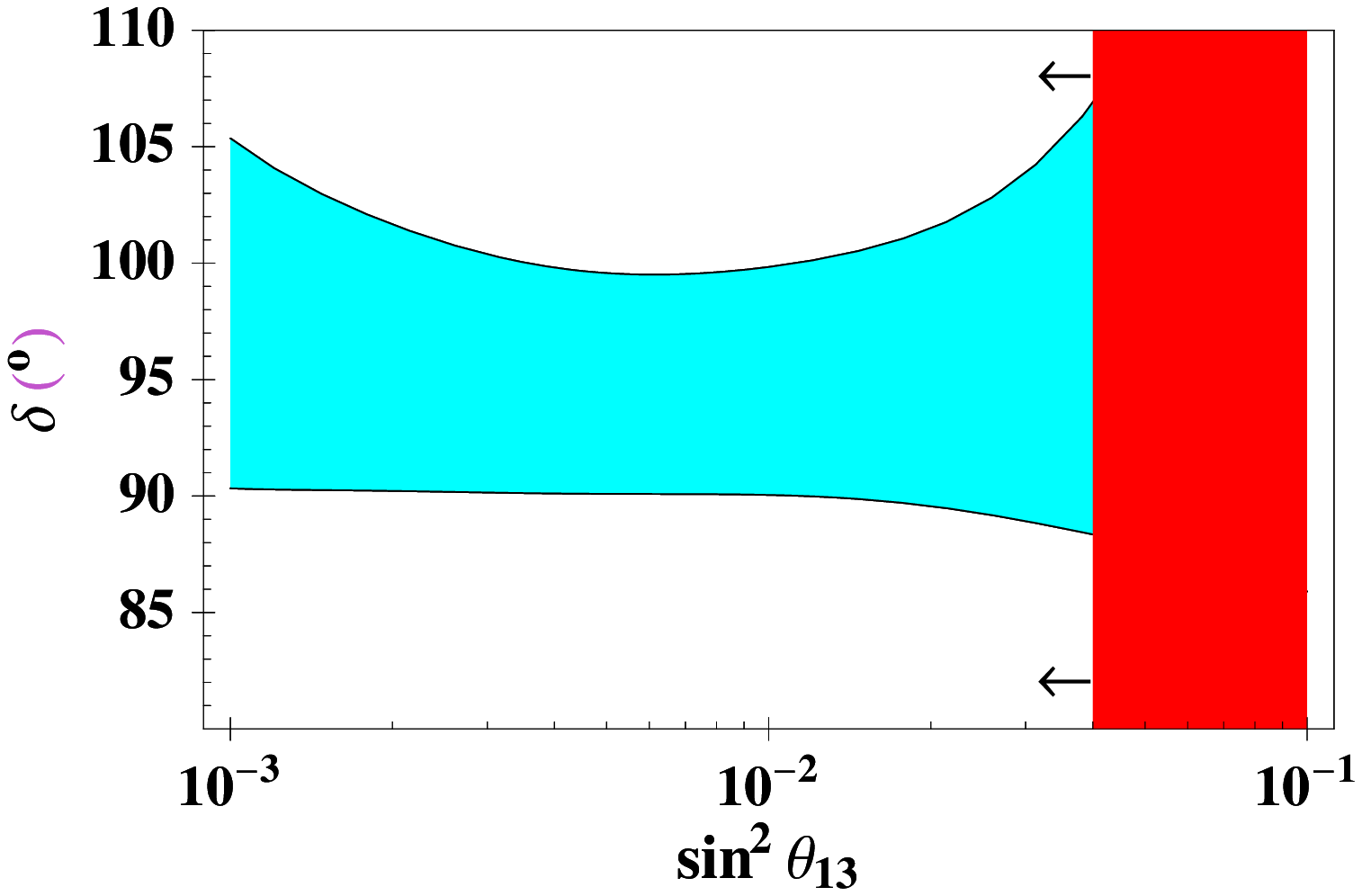}
\end{center}
\vglue -.4cm
\caption{Near-maximal CP violation in neutrino oscillations. }
\label{fig:delta}
\end{figure}
As seen in Fig.~\ref{fig:delta}, both for the $B_1$ (left panel) and $B_2$
cases (right panel) our model predicts the near maximality of the CP
violation in neutrino oscillations.
The predicted CP violating parameter $\delta$ depends mainly on
$\theta_{13}$ which is currently only bounded by oscillation
data~\cite{Maltoni:2004ei}.

\begin{figure}[h]
\begin{center}
\vglue -.3cm
\includegraphics[height=4cm,width=0.4\textwidth]{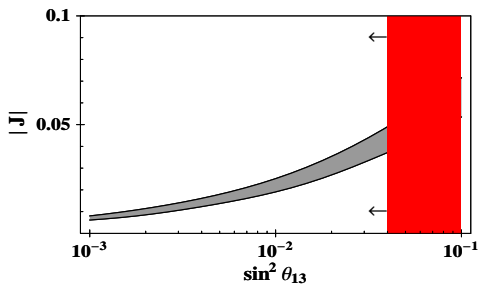} 
\end{center}
\vglue -.3cm
\caption{CP violation in neutrino oscillations versus $\sin^2\theta_{13}$. }
\label{fig:cp-inv}
\end{figure}

The rephasing invariant magnitude $|J|$ of CP violation in neutrino
oscillations is defined as 
\begin{equation}
  \label{eq:cp-inv}
  J=Im[K_{11} K_{22} K_{12}^* K_{21}^*]
  =s_{12} s_{23} s_{13} c_{12} c_{23} c_{13}^2 \sin \delta ,
\end{equation}
where $K_{ij}$ are the elements of the leptonic mixing matrix.
As seen in Fig.~\ref{fig:cp-inv}, which holds for both B1 and B2 models,
one finds that $|J|$ is directly correlated with the value of
$\sin^2\theta_{13}$, to be probed in the next generation of high
sensitivity neutrino oscillation experiments such as Double Chooz. The
width of the band reflects the current uncertainties in the neutrino
oscillation parameters~\cite{Maltoni:2004ei}.

In summary, here we have proposed an $A_4$ flavour symmetry for
leptons which leads to a near-maximal CP phase $\delta$ and correlates
the invariant measure of CP violation in neutrino oscillations with
the magnitude of $\sin^2\theta_{13}$ to be probed in future neutrino
oscillation experiments.
Moreover, it implies a lower bound $|M_{ee}| \gsim 0.03$~eV for the
mass parameter characterizing neutrinoless double beta decay, also
accessible to planned experiments. All these features already emerge
from an effective $Z_3$ invariance of the larger $A_4$ symmetry.
However, the structure of $M_R$ is different in the $A_4$ model and
the effective $Z_3$ model.  Hence, for example, some phenomenological
details related to leptogenesis could be different.  These issues will
be taken up elsewhere.
\vskip .2cm
{\bf Acknowledgments}
\vskip .2cm
Work supported by MEC grants FPA2005-01269, SAB2005-160 (A.~J) and
FPA2005-25348-E, by Generalitat Valenciana ACOMP06/154, by European
Commission Contracts MRTN-CT-2004-503369 and ILIAS/N6
RII3-CT-2004-506222.


\end{document}